\documentstyle[11pt,moriond,epsfig]{article}

\def\be{\begin{equation}}
\def\ee{\end{equation}}
\def\bea{\begin{eqnarray}}
\def\eea{\end{eqnarray}}

\def\deg{\circ}
\begin{document}
\vspace*{4cm}
\title{REVIEW OF GROUND-BASED CMB EXPERIMENTS}

\author{ \underline{A.N. LASENBY}, A.W. JONES and Y. DABROWSKI }

\address{Mullard Radio Astronomy Observatory, Madingley Road,\\
Cambridge CB3 0HE, England}

\maketitle\abstracts{
We present in this paper a brief review of Ground-Based
Cosmic Microwave Background (CMB) experiments.
We first recall the main experimental problems and adopted solutions.
We then review the Tenerife experiments, giving an update
together with some new results.
Then results and problems from other experiments are highlighted
including IAC-Bartol, Python,
Saskatoon, Mobile Anisotropy Telecope (MAT) and the
Owen Valley Radio Observatory (OVRO) experiments.
We next move on to the future ground experiments,
in particular new interferometers such as the Very Small Array (VSA),
the Cosmic Background Imager (CBI) and
the Degree Angular Scale Interferometer (DASI/VCA).
To finish, very recent work is presented on joint likelihood
analysis for estimation of cosmological parameters where both
CMB results and Large Scale Structure (LSS) surveys are considered.
}

\section{Experimental Problems and Solutions}

\subsection{The Contaminants}
From the ground, the detection of Cosmic Microwave Background (CMB) 
anisotropy at the level $\Delta T/T\sim 10^{-5}$ is a challenging
problem and
a wide range of experimental difficulties occur when conceiving and
building an experiment. We will focus here particularly on the problem
caused by the contamination of foreground sources and the solutions
that have been adopted to fight against them.
The anisotropic components that are of essential interest on angular
scales of approximately $1/2-1$ degree are: (i) The galactic dust
emission which becomes significant at high frequencies
(typically $>100$~GHz).
(ii) The galactic thermal (free-free) emission and non-thermal
(synchrotron) radiation which are significant at frequencies lower
than typically $\sim$~30~GHz.
(iii) The presence of point-like sources.
(iv) The dominating source of contamination is the atmospheric emission,
in particular at frequencies higher than $\sim$~10~GHz.

\subsection{The Solutions}
\label{sec:first}
A natural solution is to run the experiment at a suitable frequency
so that the contaminants are kept low. There exists a window
between $\sim$~10 and $\sim$~40~GHz where both atmospheric and
galactic emissions should be lower than the typical CMB flux.
For example, the Tenerife experiments are running at $10$, $15$
and $33~{\rm GHz}$ and the Cambridge Cosmic Anisotropy Telescope (CAT)
at $15~{\rm GHz}$.
However, in order to reach the level of accuracy needed,
spectral discrimination of foregrounds using multi-frequency data
is necessary (see for example M.P. Hobson, this volume).
Concerning point (iv), three basic techniques, which are all
still being used, have been developed in order to fight against the
atmospheric emission problem:

The Tenerife experiments are using the
{\it switched beam} method. 
In this case the telescope switches rapidly between
two or more beams so that a differential measurement can be made between
two different patches of the sky, allowing one to filter out the atmospheric
variations.

A more modern and flexible version of the switched beam method is the
{\it scanned beam} method (e.g. Saskatoon and Python telescopes).
These systems have a single receiver in front of which a continuously
moving mirror allows scanning of different patches of the sky.
The motion pattern of the mirror can be re-synthesised by software.
This technique provides a great flexibility regarding the angular-scale of the
observations.
The Saskatoon telescope has been very successful in using this system.

Finally, an alternative to differential measurement is the use of
interferometric techniques. Here, the output signals from each of the
baseline horns are cross-correlated so that the Fourier
coefficients of the sky are measured. In this fashion one can very
efficiently remove the atmospheric component in order to
reconstruct a cleaned temperature map of the CMB.
The Cosmic Anisotropy Telescope (CAT) operating in Cambridge has
proved this method to be very
successful, giving great expectations for the Very Small Array
(VSA) currently being built and tested in Cambridge (jointly with Jodrell Bank)
for siting in Tenerife.
American projects such as the Cosmic Background
Imager (CBI) and the Very Compact Array (VCA/DASI) are also
planning to use this technique.

\section{The Tenerife Experiments\\ {\small [Project collaborators: Rod Davies (P.I.), Carlos Guti\'errez, Bob Watson, Rafa Rebolo]}}
\subsection{Drift Scan Experiments}
Due to the stability of the atmosphere and its
transparency~\cite{davies96}, the ${\rm Iza\tilde{n}a}$ observatory
of the Tenerife island is becoming very popular for cm/mm observations
of the CMB (e.g. Tenerife experiments, IAC-Bartol, VSA).
The three Tenerife experiments ($10$, $15$ and $33~{\rm GHz}$) are
each composed of two horns using the switched beam technology
(angular resolution $\sim~5$ degrees).
The observations take advantage of the Earth rotation and consist
of scanning a band of the sky at a constant declination. The scans have
to be repeated over several days in order to achieve sufficient accuracy.
See e.g. Davies {\it et al}.~\cite{davies96} (1996) for a complete description.
 
The first detection at Tenerife (Dec+$40^{\deg}$), which dates back to
1994~\cite{hancock94,hancock97}, 
clearly reveals common structures between the three independent scans
at $10$, $15$ and $33~{\rm GHz}$.
The consistency between the three channels gave confidence that,
for the first time, identifiable individual features in the CMB were
detected~\cite{lasenby95}.
Subsequently this was confirmed by comparing directly to the COBE DMR
data~\cite{lineweaver95,hancock95}.
Bunn, Hoffman and Silk~\cite{bunn96} (1996) have applied a Wiener
filter to the COBE DMR data in order to perform a prediction for the
Tenerife experiments over the region $35^{\deg}<{\rm Dec}<45^{\deg}$.
Assuming a CDM model, the COBE angular resolution has been improved
in order to match the Tenerife experiments' resolution.
The prediction has been observationally verified~\cite{gutierrez97},
giving great confidence that the revealed features are indeed tracing
out the seed structures present in the early universe.

\subsection{Latest results}
There is now enough data to perform a 2-D sky reconstruction~\cite{aled98}
for the $10~{\rm GHz}$ and $15~{\rm GHz}$ experiments
($33~{\rm GHz}$ to follow shortly).
8 separate declination scans have been performed of the full range in RA
from Dec+$27.5^{\deg}$ up to Dec+$45^{\deg}$ in steps of $2.5^{\deg}$.
This allows the reconstruction with reasonable accuracy of a strip in the
sky of $90^{\deg}\times 17.5^{\deg}$ in an area away from major point
sources and the Galactic plane.
The crucial aspect in obtaining accurate results is, first of all, to
allow for atmospheric correlations between the different scans.
Secondly, and probably more importantly, to be aware that
the maps are sensitive to unresolved discrete radio sources
(typically of the Jy level in the Tenerife field)
in addition to the CMB. Special analysis has been performed in order
to remove these sources which have to be monitored continuously
since they are variable
on the time scales involved.
This monitoring task is done in collaboration with M. and H. Aller
(Michigan) who have a data-bank of information on these sources.

The 10~GHz 2-D map is likely to include a significant galactic
contribution; however it is believed that this contribution
is negligible
for the 15~GHz map which reveals intrinsic
Microwave Background anisotropies on 5 degrees scale.
Likelihood analysis on the reconstructed 15~GHz 2-D map is in
preparation and will be published shortly. Previous results~\cite{hancock97}
are:
$\delta T=\left[l(l+1)C_l/\left(2\pi\right)\right]^{1/2}=
34^{+15}_{-9}~{\rm\mu K}$ at $l\sim 18$ (see Table~\ref{tab:1} and
Figure~\ref{fig:ps}).
The next step concerning the data analysis is to use the Maximum
Entropy Method for frequency separation on the spherical sky,
in conjunction with all sky maps such as the Haslam 408~MHz~\cite{haslam82},
IRAS, Jodrell Bank (5~GHz) and COBE.
The resulting frequency information will allow us to
separate very accurately components such as synchrotron, free-free,
dust and CMB (see M.P. Hobson, this volume), which
is an exciting prospect.

\section{Updates and Results on Various Experiments}
\subsection{IAC-Bartol}
This experiment runs with four individual channels (91, 142, 230 and 272~GHz)
and is also located in Tenerife where the dry atmosphere is
required for such high frequencies.
This novel system is using bolometers which are coupled to
a 45~cm diameter telescope. The angular resolution is approximately
$2^{\deg}$ (see elsewhere~\cite{piccirillo91,piccirillo93,piccirillo97}
for instrument details and preliminary results).

This switched beam system has performed observations at constant
declination (Dec+$40^{\deg}$), overlapping one of the 
drift scans of the Tenerife experiments. 
Atmospheric correlation techniques between the different channels
have been applied in order to remove the strong atmospheric component
present in the three lowest channels~\cite{femenia98}.
This atmospheric contaminant is believed to be accurately removed and
the galactic synchrotron and free-free emissions are likely to be 
negligible at these frequencies. On the other hand, the galactic
dust emission has been corrected using DIRBE and COBE DMR maps.
Finally, the contamination by point-like sources was removed by
multi-frequency analysis on known and unknown sources. The results
obtained are
$\delta T=113^{+66}_{-60}~{\rm\mu K}$ at $l\sim 33$ and
$\delta T=55^{+27}_{-22}~{\rm\mu K}$ at $l\sim 53$
(see Table~\ref{tab:1} and  Figure~\ref{fig:ps}). One can notice
that the $l\sim 33$ point is well off the expected value, however,
tests show that the atmospheric component is still very high in
this $\delta T$ value. The $l\sim 53$ point seems to be in better agreement
with results from the Saskatoon or Python experiments.

\begin{table}[t]
\caption{Some Current Ground Based Experiments}
\footnotesize
\vspace{0.4cm}
\begin{center}
\begin{tabular}{c c c c c c}
\hline
Experiment & Frequency & Angular Scale & Site/Type & $l$ & $\delta T$ \\
\hline & & & & &\\
Tenerife & 10, 15, 33~GHz& $\sim 5^{\deg}$ & Tenerife (Switched Beam) &
    $18^{+9}_{-7}$ &$ 34^{+15}_{-9}$ \\ & & & & &\\

IAC-Bartol & 91, 142, & $\sim 2^{\deg}$ & Tenerife &
    $33^{+24}_{-13}$ & $113^{+66}_{-60}$ \\ 
    & 230 and 272~GHz &  & (Switched beam) &
    $53^{+22}_{-13}$ & $55^{+27}_{-22}$ \\ & & & & &\\

Python III& 90~GHz & $0.75^{\deg}$ & South Pole &
    $87^{+18}_{-38}$ & $60^{+15}_{-13}$ \\
    & & & (Scanned Beam) & $170^{+69}_{-50}$ & $66^{+17}_{-16}$ \\

Python I, II \& III& & & & $139^{+99}_{-34}$ & $63^{+15}_{-14}$ \\ & & & & &\\

    & & & & $87^{+39}_{-29}$ & $49^{+8}_{-5}$ \\ 
    & 6/12 channels & $0.5^{\deg}-3^{\deg}$ & Canada &
    $166^{+30}_{-43}$ & $69^{+7}_{-6}$ \\
Saskatoon & between & & (Scanned Beam) & $237^{+29}_{-41}$ & $85^{+10}_{-8}$\\
    & 26 and 46~GHz  &  & & $286^{+24}_{-38}$ & $86^{+12}_{-10}$ \\
    & & & & $349^{+44}_{-41}$ & $69^{+19}_{-28}$ \\ & & & & &\\

OVRO & 14.56 and 32 GHz & $\sim 0.1^{\deg}-0.4^{\deg}$ &
    Owens Valley (Switched) & $589^{+167}_{-228}$ & $56^{+8.6}_{-6.5}$ \\ & & & & &\\

 \hline
\label{tab:1}
\end{tabular}
\end{center}
\end{table}

\subsection{Python}
\begin{figure}[t]
\centerline{\epsfig{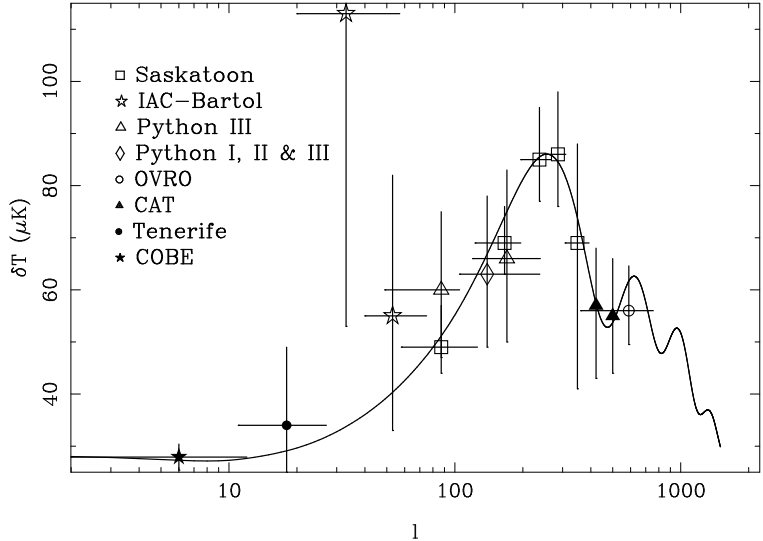}}
\caption{Plot of the CMB Power Spectrum. The data points are those
presented in Table~\ref{tab:1} and
the overlaid standard CDM curve is for the parameters estimated in
Section~\ref{sec:like}. the COBE data point is the 4-Year COBE
DMR result (Bennett~{\it et al}~$^{\rm 19}$. 1996) and the CAT points
are given in J.~Baker (this volume).}
\label{fig:ps}
\end{figure}

This experiment is using a single bolometer mounted on a 75cm
telescope and operating at the single frequency of 90GHz with a
$0.75^{\deg}$ FWHM beam.
Python is located at the Amundsen-Scott South Pole Station
in Antarctica. It is performing extremely well in terms of mapping rather
large regions of the sky (currently $22^{\deg}\times 5.5^{\deg}$).
Three seasons of observations have been analysed so far
(Python~I~\cite{dragovan94}, Python~II~\cite{ruhl95} and
Python~III~\cite{platt97}). In addition to the power-spectrum
results of Python~III (see Table~\ref{tab:1} and Figure~\ref{fig:ps}),
the combined analysis of Python~I, II \& III gives an estimate of the
power-spectrum angular spectral index~\cite{platt97}: $m=0.16^{+0.2}_{-0.18}$
which is consistent with a flat-band power model (i.e. $m=0$).

A point where the Python experiment differs from all the others
is its single frequency measurement.
All the experiments discussed here are using either
widely spaced frequencies (e.g. Tenerife experiments, COBE) or
closely patched bands of different frequencies (e.g. the interferometers
discussed in Section~\ref{sec:int}).
As mentioned above, multi-frequency analysis allows identification and
correction of the contaminanting component. However, near the pole,
the atmospheric emission is believed to be small, while at 90~GHz
the galactic dust contribution is estimated to be as small
as $\sim 2\mu K$.
On the other hand, 17 known point-like sources are present in the
Python field, which are estimated to give a 2\% effect
in the final result.
The brightest source may contribute up to $50~\mu K$ in a single
beam and ideally source removal using information from a
separate telescope at the same frequency is required.
Python~IV and V data have already been taken and the analysis
should provide power-spectrum estimations very shortly;
see Kovac~{\it et al}.~\cite{kovac97} (1997) for details about
the ${\rm IV^{th}}$ season.

\subsection{Saskatoon current status}
The Saskatoon experiment is a scanned beam system which operates
with 6 or 12 independent channels at frequencies between 26 and
46~GHz. The observations cover the North Celestial Pole
with angular scales from $0.5^{\deg}$ to $3^{\deg}$.
The experiment has been running from 1993 to 1995 
and details of the instrument as well as early results
can be found elsewhere~\cite{wollack93,wollack96}. To find more 
details about the data analysis and recent results, see for example
Wollack~{\it et al}.~\cite{wollack97} (1997),
Netterfield~{\it et al}.~\cite{netterfield97} (1997) and
Tegmark~{\it et al}.~\cite{tegmark97} (1997).

The 5 Saskatoon results (see Table~\ref{tab:1}) are crucial
in constraining the position of the first Doppler peak
(see Figure~\ref{fig:ps}) and therefore the cosmological parameters.
The overall flux calibration of the Saskatoon data was
known to have a $\pm$14\% error,
affecting significantly estimates of Hubble's parameter ($H_0$)
for example.
However, recent work from Leitch~{\it et al}. (private communication)
who carried out joint observations of Cassiopea~A and Jupiter,
allows the reduction of this uncertainty. The latest
calibration is now known with an estimated error of $\sim 4$\%.

Recent work on the foreground analysis of the Saskatoon field has
been carried out by Oliveira-Costa~{\it et al}.~\cite{oliveira97} (1997).
These authors found no significant contamination by point-like sources.
However, they report a marginal correlation between the DIRBE
$100~{\rm\mu m}$ and Saskatoon $Q$-Band maps which is likely to be
caused by galactic free-free emission.
This contamination is estimated to cause previous CMB results to be
over-estimated by a factor of 1.02.

\subsection{Mobile Anisotropy Telescope}
The Mobile Anisotropy Telescope (MAT) is using the same optics and technology
as Saskatoon at a high-altitude site in Chile (Atacama plateau at 5200m).
This site is believed to be one of the best
sites in the world for millimetre measurements and is now becoming popular
for other experiments (e.g. CBI) because of its dry weather. 
The experiment is mounted on a mobile trailer which will be towed up to
the plateau for observations and maintenance.
The relevant point where MAT differs from the Saskatoon experiment
is the presence of an extra channel operating at $140~{\rm GHz}$.
This will greatly improve the resolution and should provide results
well over the first Doppler peak. Data has already been taken over the last
few months at $140~{\rm GHz}$ and is currently being analysed.
See the MAT www-page~\cite{mathome} for a full description of the project.

\subsection{Owens Valley Radio Observatory}
The Owens Valley Radio Observatory (OVRO) telescopes have been used since
1993 for observation of the CMB at $32~{\rm GHz}$ and $14.5~{\rm GHz}$.
The RING40M experiment uses the 40-meter telescope ($14.5~{\rm GHz}$ channel)
while the RING5M experiment is mounted on the 5.5-meter telescope
($32~{\rm GHz}$ channel). Both experiments have an angular resolution 
of $\sim 0.12^{\deg}$. Details about these experiments can be found
in Readhead~{\it et al}.~\cite{readhead89} (1989) or
in Myers~{\it et al}.~\cite{myers93} (1993) for example.

36 fields at Dec+$88^{\deg}$, each separated by 22 arcmin, have been observed
around the North Galactic Pole. Using these data,
Leitch~{\it et al}.~\cite{leitch97} report 
an anomalous component of Galactic emission. Some further
work on the same data (Leitch~{\it et al},
private communication) gives the following estimate for the CMB component:
$\delta T=56^{+8.6}_{-6.5}~{\rm\mu K}$
at $l\sim 589$.
As seen on Figure~\ref{fig:ps}, this new OVRO result seems to agree well
with the CAT estimations and therefore helps in constraining the position of the
first Doppler peak. Further details of this measurement should be available
shortly.

\begin{table}[t]
\caption{Future Ground Based Experiments}
\footnotesize
\vspace{0.4cm}
\begin{center}
\begin{tabular}{c c c c c}
\hline
Experiment & Frequency & Angular Scale & Site/Type & Date \\ 
\hline
 & & & &\\

VSA  & 26 to 36~GHz & $0.25^{\deg}-2^{\deg}$ &
    Tenerife (14 element interferometer)& 1999\\
CBI  & 26 to 36~GHz & $0.07^{\deg}-0.3^{\deg}$ &
    Chile (13 element interferometer)&1999 \\
DASI & 26 to 36~GHz & $0.25^{\deg}-1.4^{\deg}$ &
    South Pole (13 element interferometer)&1999 \\
 & & & &\\ 
\hline
\label{tab:2}
\end{tabular}
\end{center}
\end{table}

\section{Ground Based Interferometers}
\label{sec:int}
As seen in Section~\ref{sec:first},
interferometers allow accurate removal of
the atmospheric component.
Therefore special ground sites are not
always necessary in order to perform sensitive measurements.
For example, the 3 element interferometer Cosmic Anisotropy Telescope
(CAT) is currently operating in Cambridge, UK~\cite{scott96}.
A full discussion of the CAT current status is given by J.~Baker (this volume)
and results, including the new CAT2 points, are plotted in Figure~\ref{fig:ps}.

The Very Small Array (VSA) is currently being built and tested in Cambridge
for siting in  Tenerife and should be observing in late 1999. The 14
elements of the interferometer will operate from 26 to 36~GHz and cover
angular scales from $0.25^{\deg}$ to $2^{\deg}$ (see Table~\ref{tab:2}).
The results will consist of 9 independent bins regularly spaced from
$l\sim 150$ to
$l\sim 900$ on the Power Spectrum diagram~\cite{jones97}.
This will give significant information on the second Doppler peak,
while the first peak will be constrained accurately enough to estimate
the total density $\Omega$ and Hubble's constant $H_0$, with a 10\% error.
The VSA current status is discussed in this volume by M.E. Jones.

There are two other interferometer projects which will complement the
work done with the VSA:
The Cosmic Background Imager (CBI), to be operated from Chile
by a CalTech team~\cite{cbi} and
The Degree Angular Scale Interferometer~\cite{white97,stark98} (DASI)
-- formerly Very Compact Array (VCA) -- which will be operated at the
South Pole (University of Chicago \& CARA).
They both share the same design (13 element interferometers)
and the same 
correlator operating from 26 to 36~GHz (see Table~\ref{tab:2}).
However the size of the baselines differ between CBI and DASI so that
CBI will cover angular scales from 4 to 20~arcmin while DASI will
cover the range between 15~arcmin and $1.4^{\deg}$ (similar to
the VSA).

All three of these interferometric experiments (VSA, CBI and DASI)
should be in operation by the end of 1999.

\section{Joining CMB and Large Scale Structures Data}
\label{sec:like}
As mentioned earlier, by comparing the observed CMB Power
Spectrum with predictions from cosmological models
one can estimate cosmological
parameters~\cite{hancock97,hancock98,lineweaver97}.
In an independent manner, similar predictions can be achieved
by comparing Large Scale Structure (LSS) surveys with cosmological
models~\cite{willick97,fisher96,heavens95}.
Recently, Webster~{\it et al.}~\cite{webster98} propose
to use full likelihood calculations in order to join CMB and LSS
predictions. They use results from various independent
CMB experiments together with the IRAS 1.2~Jy galaxy redshift survey
and parametrise a set of spatially flat models.
Because the CMB and LSS predictions are degenerate with respect
to different parameters (roughly: $\Omega_m {\rm vs}~\Omega_{\Lambda}$ for CMB;
$H_0$ and $\Omega_m {\rm vs}~b_{\rm iras}$ for LSS),
the combined data likelihood analysis
allows the authors to break these degeneracies, giving remarkable
parameter predictions. We note that $\Omega_m$ is the overall {\it matter}
density.
\begin{figure}[t]
\centerline{\epsfig{clip=,file=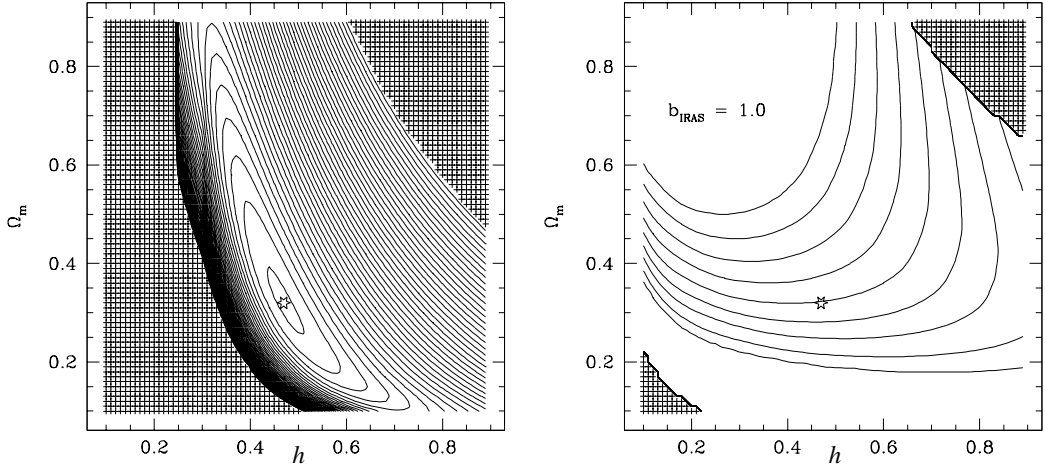,width=14cm}}
\caption{Plots on the ($\Omega_m$,$h$) plane, where
$h=H_0/100~{\rm km~s^{-1}}$. Left: Likelihood contours defining the
optimum values for ($\Omega_m$,$h$). Right: $b_{\rm iras}$ contours
with corresponding optimal value for $b_{\rm iras}=1$.}
\label{fig:cmblss}
\end{figure}
A preliminary result of the joint likelihood calculations is given in
Figure~\ref{fig:cmblss} showing a well-defined peak
for $\Omega_m$ and $H_0$, together with the corresponding
optimal value for the IRAS light-to-mass bias ($b_{\rm iras}$).

The preliminary best fit results from the two data sets joint analysis 
on all the free parameters are~\cite{webster98} (marginalised
error bars for 95\% confidence):
$\Omega_m=0.32\pm0.08$, $\Omega_b=0.061\pm0.013$,
$H_0=47\pm6$~${\rm km~s^{-1}~Mpc^{-1}}$, $b_{\rm iras}=1.04\pm0.06$ and 
$\sigma_8=0.78\pm0.1$, where $\Omega_b$ is the baryonic density
and $\sigma_8$ is the variance measured in a 8~Mpc radius sphere. 

\section{Conclusion}
Ground-based CMB experiments are already providing significant constraints
on cosmological models, particularly the interferometers and future
ground-based experiments, will sharpen these up considerably.
Although full-sky high resolution satellite experiments like
Planck Surveyor will eventually provide definite answers for the CMB,
the ability of ground-based experiments to go deep on selected patches,
and the controllability and accessibility of such experiments,
mean that they will have a very important role for some years
to come.

\end{document}